\documentclass{AGUJournal}

\journalname{JGR-Planets}

\usepackage{multirow}
\usepackage{bm}

\begin{document}

\title{Mercury's gravity, tides, and spin from MESSENGER radio science data}

\authors{Ashok Kumar Verma\affil{1} and Jean-Luc Margot\affil{1,2}}

\affiliation{1}{Department of Earth, Planetary, and Space Sciences, University of California, Los Angeles, CA 90095, USA}
\affiliation{2}{Department of Physics and Astronomy, University of California, Los Angeles, CA 90095, USA}

\correspondingauthor{Ashok Kumar Verma}{ashokverma@ucla.edu}

\begin{keypoints}
\item We estimated Mercury's gravity field, tidal Love number, and spin state based on more than 3 years of MESSENGER radio tracking data.
\item Our estimate of the Love number indicates that Mercury's mantle may be hotter and weaker than previously thought.
\item Our estimate of the spin state parameters indicates that the gravity field and crust rotate about nearly the same axis.
\end{keypoints}

\begin{abstract}
  
We analyze radio tracking data obtained during 1311 orbits of the
MESSENGER spacecraft in the period March 2011 to April 2014.  
A least-squares minimization of the residuals between observed and
computed values of two-way range and Doppler allows us to solve for a
model describing Mercury's gravity, tidal response, and spin state.
We use a spherical harmonic representation of the gravity field to
degree and order 40 and report error bars corresponding to 10 times
the formal uncertainties of the fit.  Our estimate of the product of
Mercury's mass and the gravitational constant, $GM = (22031.87404 \pm
9 \times 10^{-4})$ km$^{3}$s$^{-2}$, is in excellent agreement with
published results.  Our solution for the geophysically important
second-degree coefficients ($\bar{C}_{2,0} = -2.25100 \times 10^{-5}
\pm 1.3 \times 10^{-9}$, $\bar{C}_{2,2} = 1.24973 \times 10^{-5} \pm
1.2 \times 10^{-9}$) confirms previous estimates to better than 0.4\%
and, therefore, inferences about Mercury's moment of inertia and
interior structure.  Our estimate of the tidal Love number $k_2 =
0.464 \pm 0.023$ indicates that Mercury's mantle may be hotter and
weaker than previously thought.  Our spin state solution suggests that
gravity-based estimates of Mercury's spin axis orientation are
mar\-gin\-al\-ly consistent with previous measurements of the
orientation of the crust.
\end{abstract}

\section{Introduction}

The MErcury Surface, Space ENvironment, GEochemistry, and Ranging
(MESSENGER) mission~\citep{solo01} returned a wealth of data about the
innermost planet in the solar system.  The mission included a radio
science investigation~\citep{Srinivasan07} that provided a key
capability for characterizing Mercury's interior
structure~\citep[e.g.,][]{Smith12, maza14}.  Here we analyzed over three years
of radio tracking data with software and strategies that are different
from those used in previous investigations.

Our motivations for this investigation are fourfold: 1) The $k_2$
tidal Love number provides powerful constraints on interior models of
Mercury.  An existing determination ($k_2 = 0.451 \pm 0.014$,
\citep{maza14}) favors a cold (basal temperature of 1600 K) and stiff
(rigidity of 71 GPa) mantle with no FeS layer~\citep{pado14jgr}.
However, \citet{maza14} indicate that they cannot rule out a wider
range of values ($k_2 = 0.43-0.50$) which admit a greater variety of
plausible interior models~\citep{pado14jgr}.  We are seeking an
independent estimate of $k_2$ and its uncertainties to further
constrain interior models.  2) \citet{maza14} provided a solution for
Mercury's spin axis orientation that differs from the Earth-based
radar solution~\citep{marg12jgr}.  This may indicate an error in
either or both determinations, or a real difference between the
orientations about which the gravity field and the crust rotate.
\citet{peal16} have shown that the core spin axis may be misaligned
from the mantle spin axis, and such a difference, if present, may be
detectable.  An independent determination of the spin axis orientation
based on gravity data is needed to make progress on this issue.  3) If
the misalignment between core and mantle spin axes is sufficiently
large, the determination of the moment of inertia of the planet based
on spin and gravity data~\citep{marg12jgr} may be
jeopardized~\citep{peal16}.  In that case, the $k_2$ tidal Love number
will play an even more important role in determining the interior
structure of Mercury, and it warrants an independent determination.
4) Pre-flight simulations indicated that recovery of the longitudinal
librations should be achievable at the 8\% level from analysis of
topography and gravity data~\citep{zube97}.  Although measurements of
the librations have been obtained from Earth-based
radar~\citep{marg12jgr} and from comparison of Digital Elevation
Models (DEMs) and laser altimetry data~\citep{star15grl}, no
gravity-based estimates currently exist. Here we describe how a gravity
  solution could be used to quantify the libration signal.

Our efforts are driven by the fact that knowledge of the spin axis
orientation, Love number, and longitudinal librations are all
essential for a determination of Mercury's interior
structure~\citep[e.g.,][]{hauc13}.


\section{Spacecraft Data}
\label{scData}

Radio tracking of the MESSENGER spacecraft resulted in several
sets of observables, including one-way Doppler, two-way Doppler, and
two-way range.  

These products are available on the Planetary Data System (PDS)
\citep{tnfNote} and documentation is provided by \citet{perr11}.  The
typical precision of the two-way Doppler measurements (X-band, uplink
at 7.2 GHz, downlink at 8.4 GHz) as measured during the first 400 days
of cruise was 0.03 mm/s with an integration time of 60
s~\citep{Srinivasan07}.  \cite{maza14} showed that in-orbit residuals
were almost always less than 0.5 mm/s for Sun-Probe-Earth (SPE) angles
exceeding 40 degrees.  

{
Two-way Doppler relies on coherent repetition
of a signal by an on-board transponder and circumvents frequency drift
problems associated with on-board oscillators.
Because one-way
Doppler tracking is less accurate than two-way Doppler tracking, and
because it represents a small fraction ($<$6\%) of the entire data, we
did not include one-way data in our analysis.  Three-way data were not
used in the analysis.}

MESSENGER communications with Earth relied on the Deep Space Network
(DSN) stations located in Goldstone (California), Madrid (Spain), and
Canberra (Australia) and 8 non-steerable antennas mounted on the
spacecraft. Two phased-array high-gain antennas (hereafter PAAs)
provided the highest quality link.
Two fanbeam antennas collocated with the PAAs provided a medium-gain
link. The spacecraft was also equipped with 4 low-gain antennas
(LGAs)~\citep[][]{Srinivasan07}.
MESSENGER operations relied on various combinations of these antennas
depending on orbital geometry, spacecraft attitude, time of day, and
scheduling constraints.

Earth-based antenna locations are known with an accuracy of
centimeters in the 1993 realization of the IERS Terrestrial Reference
Frame \citep{Folkner97,DSN301J}. Spacecraft antenna coordinates are
available in PDS documents \citep{scAntNote}
and listed with 5 digits in Table~\ref{tab-scAnt}.  
We applied the 0.89662 m correction to
the spacecraft antenna coordinates~\citep{scAntNote}.

\begin{table}[h]
\caption{Antenna offsets$^{a}$ in the spacecraft frame with origin at the adapter ring.  LGA: low-gain antenna; FBA: fanbeam antenna; PAA: phased-array antenna.}
\centering
\begin{tabular}{ l c c c c}
\hline  
    Antenna       &    Type   &  X(m) & Y(m) & Z(m) \\ 
\hline
Front LGA & low-gain    & -0.1270   &  -1.0348 &  -1.9939  \\
Back LGA  & low-gain    &  0.1095   &   1.2753 &  -1.4262  \\
AFT LGA   & low-gain    & -0.2794   &  -0.8593 &   0.3686  \\
FWD LGA   & low-gain    &  0.2794   &  -0.8593 &  -2.2291 \\
Front FBA & medium-gain & -0.7272   &  -0.6487 &  -1.6162 \\
Back FBA  & medium-gain &  0.2743   &   1.1394 &  -1.2603 \\
Front PAA & high-gain   & -0.7272   &  -0.6487 &  -1.6162 \\
Back PAA  & high-gain   &  0.2743   &   1.1394 &  -1.2603 \\
\hline
\multicolumn{5}{l}{$^{a}$MESSENGER center of mass is offset by 0.89662 m} \\
\multicolumn{5}{l}{ from the adapter ring along the Z direction} \\
\multicolumn{5}{l}{\citep{scAntNote}.}
\label{tab-scAnt}
\end{tabular}
\end{table}

Tracking of the MESSENGER spacecraft by Earth-based antennas was not
continuous.  During the first (January 14, 2008) and third (September
29, 2009) flybys of Mercury, MESSENGER was occulted by Mercury
immediately before and after the closest approach, respectively.  No
radio link was possible during the occultation periods.  MESSENGER was
tracked continuously during the second Mercury flyby (October 6,
2008).  During the orbital phase of the mission, MESSENGER was tracked
for about 8 hours/day, except in the first couple of weeks of orbital
operations where it was tracked for about 16 hours/day.  After the
2012 mission extension, when MESSENGER's orbital period was reduced
from 12 hours to 8 hours, MESSENGER was tracked for about 6 hours/day.

\section{Force and Measurement Models}
\label{models}

We used the Mission Operations and Navigation Toolkit Environment
(MONTE) software \replaced {[Flangan and Ely, 2012]}{\citep{Evans16}} for orbit determination and
parameter estimation.  MONTE is developed and maintained by NASA's Jet
Propulsion Laboratory (JPL). MONTE allows for precision modeling of
the forces that act on the spacecraft and of the observables.

MONTE numerically integrates the equations of motion and computes
partial derivatives with respect to the solve-for parameters.  The
MONTE integrator uses a variable-order Adams method for solving
ordinary differential equations and is well-suited for integrating
trajectories.

A MONTE filter minimizes the residuals by adjusting the solve-for
parameters and computes parameter uncertainties.
The process relies on a UD-Kalman filter~\citep{Bierman77}, where $U$
is an upper triangular matrix with the diagonal elements equal to one,
and $D$ is a diagonal matrix.

\subsection{Gravitational Force Modeling}
\label{gfm}
MONTE's representation of the gravity field follows the traditional
spherical harmonic expansion of the
potential~\citep[e.g.,][]{Kaula00}:

\begin{equation}
 \label{potential}
U =  \frac{GM}{r} + \frac{GM}{r} \sum_{l=2}^{\infty}\sum_{m=0}^{l} \left(\frac{R}{r}\right)^l \bar{P}_{l,m} (sin \ \phi) \nonumber 
   \bigg(\bar{C}_{l,m} \ cos(m\lambda) + \bar{S}_{l,m} \ sin(m\lambda) \bigg), 
\end{equation}
\noindent where $G$ is the gravitational constant, $M$ is the mass of
Mercury, $\bar{P}_{l,m}$ are the normalized associated Legendre
polynomials of degree $l$ and order $m$, $R$ is the reference radius
of Mercury (2440 km), and $\lambda$, $\phi$, and $r$ are the
planetocentric longitude, latitude, and distance of MESSENGER from the
origin of the reference frame, which is chosen to coincide with
Mercury's center of mass.  $\bar{C}_{l,m}$ and $\bar{S}_{l,m}$ are the
normalized dimensionless spherical harmonic coefficients.  We used the
full normalization as described in \citet[][p.7]{Kaula00}.

In addition to Mercury's gravitational forces, other gravitational forces included in
our force budget are: relativistic perturbations; gravitational perturbations from 
the Sun, Earth, and other planets computed from the DE432 
\citep{FolknerDE432} planetary ephemerides;  and perturbations due to solid tides 
raised on Mercury by the Sun.

We used the Earth-based radar solution \citep{marg12jgr} as an priori
estimate of Mercury's spin axis orientation, and we used the IAU value
of the mean resonant spin rate~\citep{arch11}.  Our libration model is
based on the formulation of \citet{marg09cmda}, where all amplitude
coefficients have been scaled by a factor of 1.074 to account for
improved estimates of the libration amplitudes between the initial
~\citep{marg07} and more recent ~\citep{marg12jgr} estimates.

\subsection{Non-Gravitational Force Modeling}
\label{ngfm}
The main non-gravitational perturbations that affect MESSENGER's
trajectory are solar radiation pressure, sunlight reflected by
Mercury, thermal radiation emitted by Mercury, and propulsive
maneuvers.

We used an eleven-element box model \citep{Vaughan02} to compute
radiative forces on the MESSENGER spacecraft.  In this model, one
cylindrical element represents the spacecraft bus, eight flat plates
represent the spacecraft sunshade, and two flat plates represent the
two solar panels.  The contribution of each individual element to the
spacecraft acceleration was computed on the basis of surface area,
specular and diffuse reflectivity parameters, and the element's
orientation in the body-fixed frame of the spacecraft.
The orientation of the spacecraft bus and of the articulated solar
panels can be obtained from quaternions defined in attitude data
kernels provided by the MESSENGER
team \citep{ckNote}.

The contributions of individual elements were then summed vectorially
to obtain the total non-gravitational acceleration due to the Sun
(direct solar radiation) and Mercury (albedo and thermal emissivity).
We assumed a uniform albedo distribution (0.074) and surface
temperature for Mercury.  The magnitude of the radiation pressure due
to Mercury is an order of magnitude smaller than the solar radiation
pressure. Accurate modeling of radiation pressure forces depends on
the physical properties of the spacecraft box model, which are not
perfectly known. In order to account for the inevitable mismodeling of
these forces, we adjusted parameters representing the area of the
model elements, the specular and diffusive reflectivity coefficients,
and three scale factors (Section \ref{solvePara}).  
This approach
takes care of modeling errors and is robust against the details of the
spacecraft box model.

\subsection{Measurement Modeling}
\label{mm}

Two-way radiometric measurements encode the propagation of the radio
signal from the Earth-based antenna to the spacecraft antenna,
coherent turnaround of the signal, and propagation back to the ground
station.  We extracted the observables (range and Doppler) from the
Tracking and Navigation Files (TNF) \citep{tnfNote} according to
specifications of the TRK-2-34 format~\citep{trk34}.  We compressed
the Doppler data by using an integration time of 30~s.  
{Over the
course of 30~s, the spacecraft traveled distances of at most $\sim$120
km, which is less than the resolution of our expansion of the gravity
field to degree and order 40.}

We applied media corrections to the raw radiometric data to correct
for the effects of the Earth's troposphere \citep{troNote} and ionosphere \citep{inoNote}.  Our
tropospheric refraction delay model \citep{Moyer, Estefan94} is
composed of wet and dry mapping functions and their respective wet and
dry zenith delays \citep{Niell96}. We used meteorological data \citep{metNote}
collected at each DSN sites to compute these propagation delays.

Modeling of the DSN station positions included the effects of Earth's
precession and nutation, Earth's polar motion, solid-Earth tides,
ocean loading, and tectonic plate motion~\citep{DSN301J}.  The
positions of the spacecraft antennas with respect to MESSENGER's
center of mass (Table~\ref{tab-scAnt}) were held fixed.

Our processing of the Doppler and range data used the formulations by
\citet{Moyer} which are implemented in MONTE.  Our propagation model
includes a spacecraft transponder delay of 1379 ns~\citep{Srinivasan07}.

\section{Solution Technique}
\label{solTq}

Our objectives were to obtain an independent solution for Mercury's
gravity field, spin axis orientation, and Love number $k_2$. In this
section we describe our procedures for determining the orbit of
MESSENGER and for retrieving the geophysical parameters of Mercury.

\subsection{Data Extent}
\label{daDis}

We analyzed MESSENGER tracking data from the Mercury flybys that
provided good radio science coverage (flyby 1 and 2).  We also
analyzed over 3 years of data from the orbital phase of the
mission, from March 2011 to April 2014.  With our integration time of
30 s, this data set contains about 780,000 Doppler points and 35,000
range points.  The flyby data are important because the spacecraft was
following a roughly East-West trajectory over the equator at the times
of closest approach to the planet ($\sim 200$ km above the surface),
whereas the spacecraft followed a roughly North-South trajectory and
remained at higher altitudes when crossing the equator ($> 1000$ km
above the surface) after orbit insertion.

We divided the radiometric data from the orbital phase into 2963 arcs,
where each arc represents a continuous span of time
corresponding to MESSENGER's orbital period.  We selected the
beginning of each arc to correspond to MESSENGER's apoapsis, i.e., the
time when MESSENGER reached its farthest distance from Mercury.  For
the two flyby arcs that we processed, we selected the arc length
corresponding to the interval during which MESSENGER was flying within
Mercury's sphere of influence~\citep[][p. 352--353]{danb03},
approximately 10 and 11 h, respectively.

Several hundred arcs contained no radio science data at all because
arc lengths are 8 to 12 hours and DSN tracking typically lasted 6 to 8
hours per day.  Empty arcs were discarded.

We eliminated 914 arcs with Sun-Probe-Earth (SPE) angles below $< 40$
degrees.  These arcs occur near superior solar conjunctions, where
strong turbulent and ionized gases within the solar corona severely
degrade the radio wave signal \citep{asma05}.  This degradation causes
phase delays in the signal which are directly proportional to the
total electron density and inversely proportional to the square of
carrier radio wave frequency \citep{asma05,verma13}.

We also excluded 141 arcs that were affected by spacecraft
maneuvers, including momentum dump maneuvers (MDM) and orbit
correction maneuvers (OCM).
Doing so allowed us to avoid errors due to strong unmodeled or
mismodeled accelerations.

Finally, we excluded 4 arcs for which attitude information was
incomplete and 9 arcs for which inspection of the pattern of residuals
revealed obvious problems in data collection or quality.
Among the remaining arcs, we excluded a number of outliers, i.e.,
individual data points that had unusually large residuals compared to
simulated observables computed on the basis of a reference trajectory
(Doppler residuals in excess of 100~mHz and range residuals in excess
of 1000~m).  For this step, we used the trajectory produced by the
navigation team, which was expedient but not required.  The same
outlier rejection step could have been performed with our own
reconstructed trajectory, albeit with some additional processing. The
fraction of outliers was about 3.1\% of all tracking data.
After this process of elimination, 1311 arcs remained.

\subsection{Batch Processing}
We distributed all 1311 arcs into 10 batches according to a simple
prescription: batch $i$ ($0 \leq i \leq 9$) contains all arcs with an
arc number ending in $i$.  The number of arcs per batch ranges from
124 to 135.  This distribution resulted in a roughly uniform sampling
of the geometrical circumstances and provided thorough longitudinal
coverage in each batch.

The computational cost of the MONTE filter scales roughly as $N^2M$,
where $N$ is the number of solve-for parameters and $M$ is the number
of data points.  Splitting the arcs into 10 separate batches enabled
processing in parallel (one batch per computing node) and reduced the
computational time by about an order of magnitude.  This reduction in
computational cost is the primary advantage of this batch processing
technique.

\replaced{A potential disadvantage of splitting the data into batches as opposed 
to inverting all the data at once is that one or more batches could yield anomalous 
parameter solutions. We verified that there is excellent consistency across the 10 
batch solutions by computing the corresponding standard deviations (Sections 5.3 and 5.4).)}

{In the weighted least-squares estimation technique used in many
geodesy applications, the entire data set is traditionally processed
as a single batch~\citep[e.g.,][]{tapl04,mont12}.  Our approach
involves distributing the data in 10 separate batches.
A potential disadvantage of splitting the data into batches as opposed
to inverting all the data at once is that one or more batches could
yield anomalous parameter solutions.  We verified that there is
excellent consistency across the 10 batch solutions by computing the
corresponding standard deviations (Sections~\ref{loveK2} and
\ref{spinAxis}).  Another potential disadvantage of using multiple
batches is that the traditional covariance matrix cannot be readily
obtained.  However, it is possible to approximate this matrix by
taking the average of the covariance matrices obtained in each one of
the 10 batches.}
 
\subsection{Solve-for Parameters}
\label{solvePara}
We used MONTE's filter module to refine estimates of solve-for
parameters and their formal uncertainties.  We grouped the estimated
parameters into 3 categories: {\em arc-level}, {\em batch-level},
and {\em global}.

{\em Arc-level} parameters are those parameters that affect the
measurements within a single arc.  They consist of the spacecraft
state vector (6 parameters) at the initial epoch of each arc and of 3
scale factors to account for the mismodeling of accelerations due to
solar radiation pressure, sunlight reflected by Mercury, and thermal
emission from Mercury.  The number of {\em arc-level} parameters is
therefore equal to 9 times the number of arcs processed.

{\em Batch-level} parameters are those parameters that affect the
measurements across multiple arcs.  They consist of the $GM$ of
Mercury, 1677 coefficients for a spherical harmonic expansion to
degree 40, the tidal Love number $k_2$, 2 angles specifying the spin
axis orientation of Mercury, 3 parameters specifying the position of
the spacecraft center of mass with respect to the spacecraft reference
frame, 3 parameters for each one of the 11 elements of the spacecraft
box model (surface area, specular reflectivity coefficient, diffuse
reflectivity coefficient), and
Doppler and range biases for each one of the 12 ground-based antennas.
{Doppler and range biases are designed to absorb small errors in the
knowledge of antenna positions or other undetected systematic effects.
The Doppler biases ($<3$ mHz) are smaller than the RMS values of the
residuals (Section~\ref{gravField}).  The range biases capture any
bias in the knowledge of the Earth-Mercury distance (Section~\ref{merEph}).}
The total number of {\em batch-level} parameters is 1741.

{\em Global} parameters are a subset of {\em batch-level} parameters
that have geophysical significance.  They consist of the $GM$ of
Mercury, spherical harmonics coefficients, the tidal Love number
$k_2$, and two angles specifying the spin axis orientation of Mercury.
{\em Global} parameters were calculated by taking the weighted mean of the
independent estimates obtained from the 10 batches (Section
\ref{solSta}).

\subsection{Solution Strategy}
\label{solSta}

Our strategy consisted of 3 successive steps: numerical integration
of the spacecraft state and its partial derivatives with respect to
solve-for parameters, computation of range and Doppler values and
their partial derivatives with respect to solve-for parameters at the
epochs corresponding to the observables, and optimization of the {\em
  arc-level} and {\em batch-level} parameters to minimize the
residuals between observed and computed values.

\begin{figure}[h]
  	\centering
\noindent
\includegraphics[width=30pc]{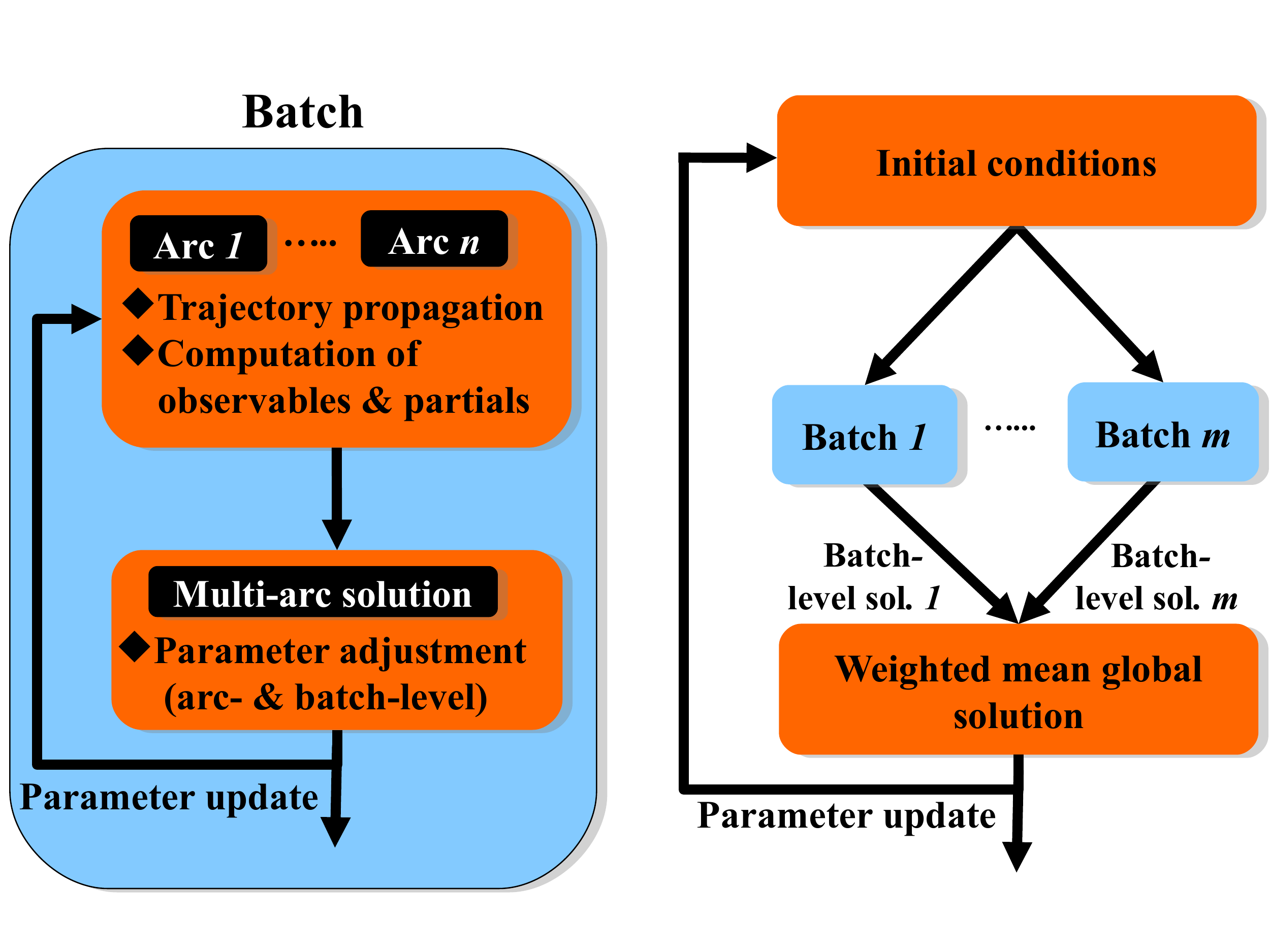}
\caption{Orbit solution strategy.}
\label{fig-solTecq}  
\end{figure}

Figure \ref{fig-solTecq} illustrates our solution strategy.
Batch $i$ ($1
\leq i \leq 10$) includes $n_i$ data arcs.
As an expedient but optional step, we assigned the position and
velocity of MESSENGER from the navigation
solution \citep{spkNote}
as the initial state vector at the start of each arc.  We performed
separate orbit integrations for each arc and did not attempt to link
integrations from consecutive arcs.  An alternate choice would have
been to use the state vector at the end of one arc to initiate the
orbit integration for the next arc.

After integration of the orbit with the current values of the {\em
  arc-level} and {\em batch-level} solve-for parameters, we used our
observation model (Section \ref{mm}) to compute range and Doppler
values at the epochs corresponding to the observables.  We computed the
root-mean-squared (RMS) values of the pre-fit residuals as 
\begin{equation}
       {\rm RMS}_j = \sqrt{\frac{1}{m_j}\sum_{k=1}^{m_j}(O_k-C_k)^2},
       \label{eq-rms-arc}
\end{equation}
where $m_j$ is the number of data points in arc $j$ ($1 \leq j \leq
n_i$), $O_k$ is the k-th observable, and $C_k$ is the corresponding
computed value.  We used the RMS values as an indication of the data
quality in each arc and assigned observational uncertainties
corresponding to ${\rm RMS}_j$ to each data point in arc $j$.
{
Because most of the low-altitude data were acquired with the LGAs and
because these data provide the best leverage on the recovery of the
gravity field, we found that it was more effective to assign uniform
uncertainties than to assign uncertainties on the basis of antenna
gain.}

Once the computed observables were generated for all arcs in a batch,
we used MONTE's optimization filter {(Section~\ref{models})} to
simultaneously adjust {\em arc-level} and {\em batch-level}
parameters.  This optimization was informed by a priori uncertainties
for some of the parameters (Section~\ref{modEst}).  For the spin axis
orientation and $k_2$ Love number, we used a bounded \added{($\pm 3$ times a
priori uncertainty)} optimization technique during the first iteration
to rule out implausible values.  The batch-level adjustments to the
parameters is an iterative process resulting in new integrations with
the updated parameter values (Fig.~\ref{fig-solTecq}, left).  We
stopped iterating when the change in the RMS value of the post-fit
residuals for the entire batch
\begin{equation}
           {\rm RMS}_i = \sqrt{\frac{1}{n_i}\frac{1}{m_j}\sum_{j=1}^{n_i}\sum_{k=1}^{m_j}(O_k-C_k)^2},
       \label{eq-rms-batch}
\end{equation}
decreased by less than 10$\%$ compared to the previous iteration.
{
Because batch-level estimates are further combined and used in a
global iterative process with 10 iterations, we found that it was not
worthwhile to decrease the batch-level stopping criterion below 10\%.}

Each batch provided an independent estimate of each one of the {\em
  global} solve-for parameters.  To combine these estimates, we used a
weighted mean technique in which a parameter estimate and its variance
are given by:
\begin{equation}
 \label{wgtMean2}
           p = \frac{\sum_{i=1}^{10} w_i p_i}{\sum_{i=1}^{10} w_i}
\end{equation}
\begin{equation}
 \label{wgtMean3}
           \sigma_p^2 = \frac{1}{\sum_{i=1}^{10} w_i}
\end{equation}
where $w_i=1/{\sigma_{p,i}^2}$ is the weight corresponding to batch
$i$ and $\sigma_{p,i}$ is the formal uncertainty associated with
parameter $p$ and batch $i$.

We used the weighted mean estimates and their uncertainties to update
the dynamical model and iterated the entire process
(Fig.~\ref{fig-solTecq}, right).  \added{Arc-level and non-global parameters
were reset to their nominal values.}  After 10 iterations, the
high-order gravity coefficients had an error spectrum that stabilized
near the Kaula constraint (Section~\ref{gravField}) and other
parameters exhibited variations that did not exceed the formal
uncertainties of the fit.


\section{Model Estimation}
\label{modEst}

Our model estimation consisted of two major steps: analysis of
flyby data (Section \ref{flyby}) followed by estimation of a gravity
field to degree and order 40 on the basis of both flyby and orbital
data (Section \ref{gravField}).  We discuss separately our estimate of
the tidal Love number $k_2$ (Section \ref{loveK2}), spin axis
orientation (Section \ref{spinAxis}), 
{and Earth-Mercury distance error (Section \ref{merEph})}. 

\subsection{Flyby Analysis}
\label{flyby}

We were interested in validating our ability to recover certain
gravity quantities using only flyby data, and, therefore, we first analyzed
MESSENGER flyby data as if no orbital data existed.

During the first and third
flybys, Mercury occulted MESSENGER, preventing radio frequency
transmission near closest approach.  Moreover, MESSENGER went into
safe mode just before its closest approach to Mercury on the third
flyby, resulting in the loss of tracking data.  For these reasons, we
did not include the third flyby in our analysis.

 In our first estimation, we concentrated on the values of $GM$,
$\bar{C}_{2,0}$, and $\bar{C}_{2,2}$.  We used the Mariner 10
estimates of these quantities \citep{Anderson87} as a priori estimates
for this initial flyby solution.  Other gravity coefficients were held
fixed at zero.  We verified that our procedures converged on the
correct $GM$ and $\bar{C}_{2,2}$ values even if we provided a priori
information that differed markedly from the Mariner 10 values (e.g.,
$GM=1000$ km$^{3}$s$^{-2}$, $\bar{C}_{2,2}=0$).  After this initial
step, we used our estimates of $GM$, $\bar{C}_{2,0}$, and
$\bar{C}_{2,2}$ as a priori estimates for a gravity solution to degree
and order 4.

Our second estimation consisted of a gravity solution to degree and
order 4 (HgMUCLA4x4).  There was not sufficient coverage to expect a
reliable recovery of $\bar{C}_{2,1}$ and $\bar{S}_{2,1}$, so these
coefficients were held fixed at zero.  All the gravity coefficients
from this solution are within error bars of the coefficients obtained
in the HgM001 flyby analysis of \citet{smit10}.  Table~\ref{tab-flybySol}
shows the most important values extracted from the HgMUCLA4x4 and
HgM001 solutions.
\begin{table}[h]
\caption{Comparison of flyby data solutions HgMUCLA4x4 (this work) and
  HGM001 (\citet{smit10}), with formal uncertainties of the fit.}
\centering  
\begin{tabular}{ l r r}
\hline  
    Parameter       &    HgMUCLA4x4 & HgM001 \\ 
\hline
GM (km$^{3}$s$^{-2}$)  & 22031.88 $\pm$ 0.19 &  22031.80 $\pm$ 0.08\\
$\bar{C}_{2,0}$ (10$^{-5}$) & -0.171 $\pm$ 0.5 &  -0.857 $\pm$ 1.8 \\
$\bar{C}_{2,2}$ (10$^{-5}$)  & 1.252 $\pm$ 0.7 &    1.258 $\pm$ 0.7 \\
\hline
\label{tab-flybySol}
\end{tabular}
\end{table}

The equatorial geometry of the MESSENGER flybys provided good
sensitivity to the dynamical equatorial flattening $\bar{C}_{2,2}$ but
poor sensitivity to the dynamical polar flattening $\bar{C}_{2,0}$.
We therefore regard our initial estimate of $\bar{C}_{2,0}$ as
unreliable, as did \citet{smit10}.  Nevertheless, the $GM$ and
$\bar{C}_{2,2}$ values are both fully consistent with the analysis of
orbital data. 
{We used the HgMUCLA4x4 estimates of $GM$,
$\bar{C}_{2,0}$, and $\bar{C}_{2,2}$ as a priori estimates for a
gravity solution to degree and order 40.}

\subsection{Gravity Field Solution HgMUCLA40x40}
\label{gravField}

\begin{figure}[h]
\centering
\includegraphics[width=\textwidth]{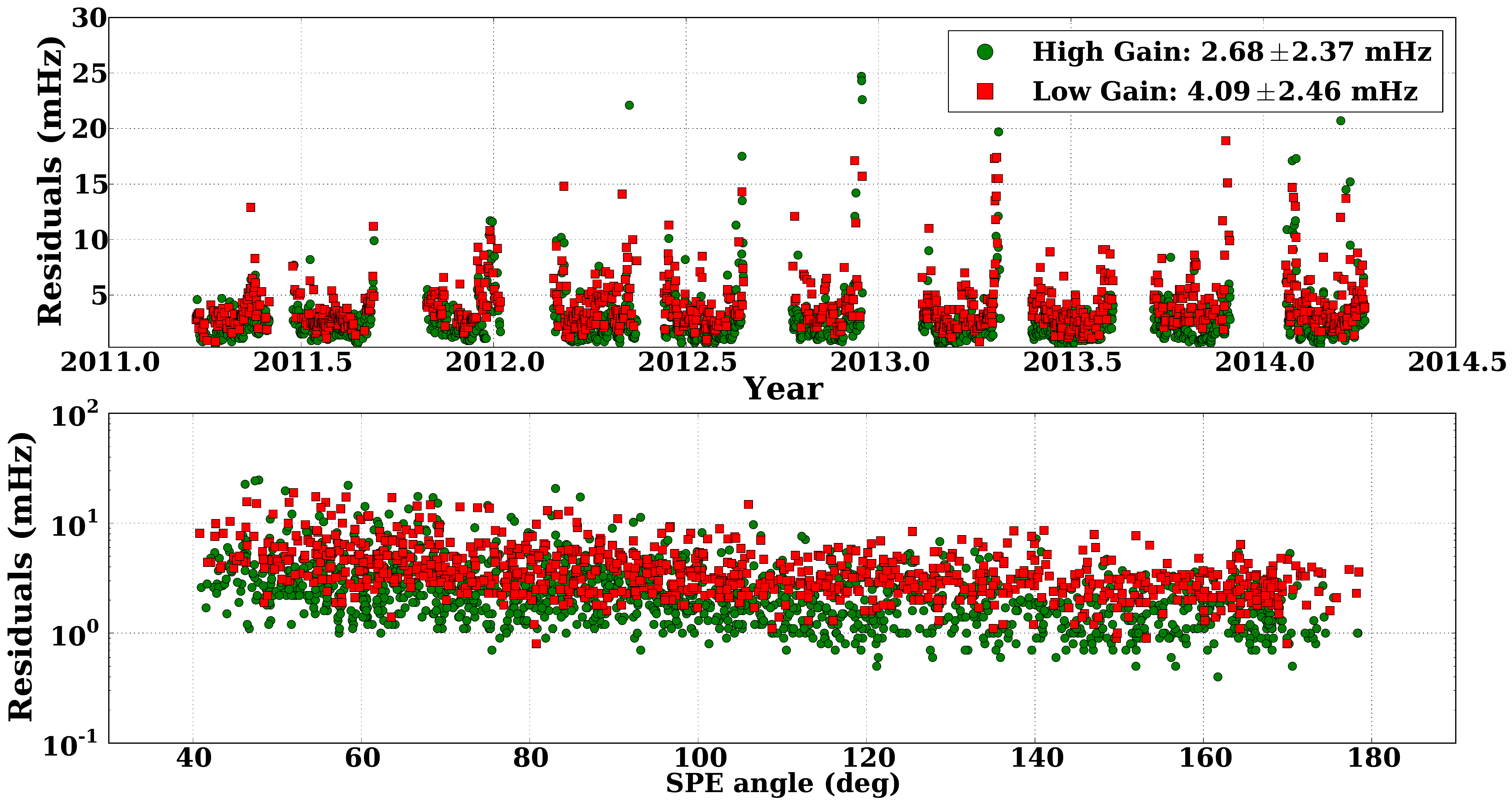}
\caption{RMS values of two-way Doppler residuals obtained after
  fitting arc-level parameters for each data arc using the HgMUCLA40x40
  gravity field solution, shown as a function of time (Top) and SPE
  angle (Bottom).  The visible gaps between clusters of points
  correspond to periods of superior conjunction (SPE $\leq 40^\circ$),
  which we excluded from the analysis (Section \ref{daDis}).  Each
  point in the figure represents the RMS value of residuals for
  individual arcs as computed by equation (\ref{eq-rms-arc}).
  Residuals associated with high-gain antennas are generally lower
  than those associated with low-gain antennas, as expected.  Values
  in the rectangular box represent the mean and standard deviation of
  the residuals calculated over the entire data set.}
\label{fig-residuals}
\end{figure}

The gravity field solution that we obtained from more than 3 years of
MESSENGER tracking data, hereafter HgMUCLA40x40, produces very good
measurement residuals (Fig.~\ref{fig-residuals}).  
The RMS values of the residuals are better than the instrument measurement requirement
\citep[$\sim$0.1 mm/s or $\sim$5.6 mHz,][]{Srinivasan07}
and give confidence about the quality of the fit and the recovery of
estimated parameters.

\begin{figure}
\centering
\includegraphics[width=25pc]{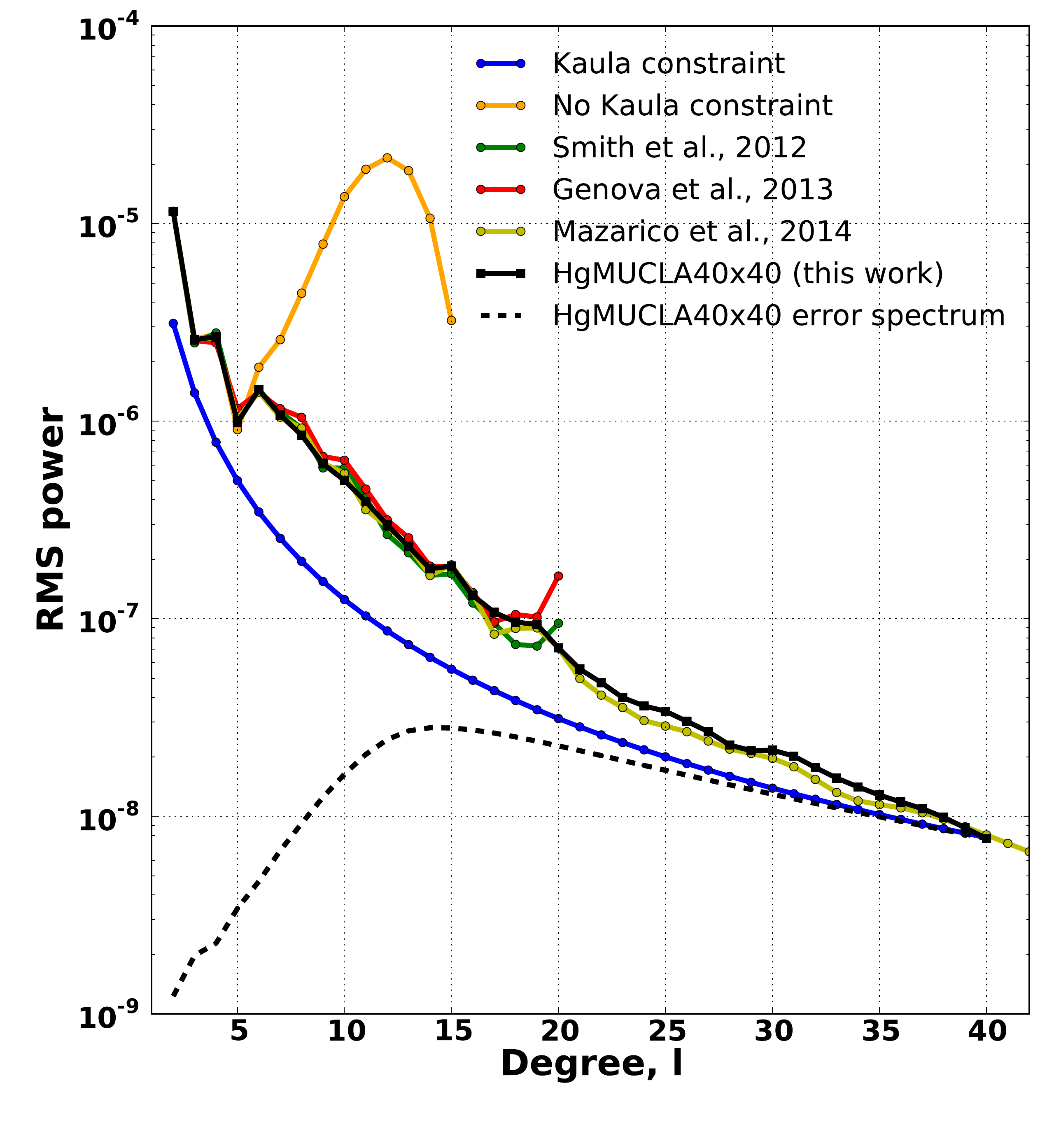}
\caption{Power associated with degree $l$ in several spherical
  harmonic expansions of the gravity field. HgMUCLA40x40 (this work)
  is shown with the solid black line.  The solid blue line illustrates
  the a priori Kaula constraint that was used for degrees $l > 6$.
  The dashed black line illustrates the error spectrum.  The magnitude
  of the formal uncertainties associated with our analysis approaches
  the magnitude of the coefficients at degree $l \sim 40$, where the
  solid and dashed black lines intersect.  The solid orange line
  depicts the power spectrum obtained when the Kaula constraint is not
  applied.}
\label{fig-gravRMSpower}
\end{figure}

The power associated with degree $l$ in a spherical harmonic expansion
is given by
\begin{equation}
 \label{rmsPower}
           P_l = \sqrt{\frac{1}{(2l + 1)}  \sum_{m=0}^{l} (\bar{C}_{l,m}^{2} +  \bar{S}_{l,m}^{2} )}.
\end{equation}
{The power associated with HgMUCLA40x40 harmonics of degree
$l$ compares favorably with that of other solutions
(Fig.~\ref{fig-gravRMSpower}).}
In order to limit the power at high degrees in our solution to degree
and order 40, we placed a priori constraints on the uncertainties
associated with each coefficient of degree $l$
($7 \leq l \leq 40$) equal to $K/l^2$. This is the well-know Kaula
constraint.  We experimented with a few values of $K$ and settled on
the value adopted by \citet{maza14}, $K = 1.25 \times 10^{-5}$.
Application of a Kaula constraint helped mitigate against spurious
results due to the lack of low-altitude coverage in the southern
hemisphere, such as the development of large gravity anomalies.
{As a test, we
obtained a gravity solution to degree and order 15 without applying
the Kaula constraint to verify that this constraint is necessary for
degrees $l > 6$ (Fig.~\ref{fig-gravRMSpower}).}

The HgMUCLA40x40 values of $GM$ and low-degree coefficients
(Table~\ref{tab-gravSol40}) are in excellent agreement with the HgM005
solution of \citet{maza14}.  The low-degree coefficients play a vital
role in understanding the planet's interior
structure~\citep[e.g.,][]{marg12jgr}.  In particular,
the second-degree coefficients ${C}_{2,0}$ and ${C}_{2,2}$, combined
with spin state parameters, can be used to estimate the moment of
inertia of the planet and that of its core.
We find that ${C}_{2,0} = -5.033 \times 10^{-5}$ and ${C}_{2,2}
  = 8.067 \times 10^{-6}$ are within 0.05\% and 0.3\% of the
  \citet{Smith12} values that have been used to infer properties of
  Mercury's interior~\citep{marg12jgr,hauc13,rivo13}, adding
  confidence to these studies.  Mercury's oblateness $J_2 =
  -{C}_{2,0}$.

\begin{table}
  \footnotesize
  \setlength{\tabcolsep}{4pt}
  \caption{Select coefficients of HgMUCLA40x40 solution (this work) compared with those of the HgM005 solution \citep{maza14}, showing the fractional change $\Delta$ between the solutions.  Our adopted error bars are 10 times the formal uncertainties listed.}
\centering
\begin{tabular}{lrrrrrrrrr}\hline
\multicolumn{1}{c}{\multirow{2}{*}{{\bf Parameter}}} &
\multicolumn{4}{c}{{\bf HgMUCLA40x40}}     & 
\multicolumn{2}{c}{{\bf HgM005}} &\multicolumn{1}{c}{\multirow{2}{*}{$\boldsymbol{\Delta}$}} &&
\\ \cline{3-4} \cline{6-7} && 
\multicolumn{1}{c}{\bf Value } & \multicolumn{1}{c}{\bf Formal Uncert.} &&
\multicolumn{1}{c}{\bf Value } & \multicolumn{1}{c}{\bf Formal Uncert.} \\
\hline
GM (km$^{3}$s$^{-2}$)        && 22031.87404                    &  9 $\times$ 10$^{-5}$  &&  22031.87080                     &    9 $\times$ 10$^{-5}$ & -1.5 $\times$ 10$^{-7}$ \\
$\bar{C}_{2,0}$              && -2.25100  $\times$ 10$^{-5}$   &  1.3 $\times$ 10$^{-9}$ &&   -2.25045  $\times$ 10$^{-5}$   &  0.7 $\times$ 10$^{-9}$ & -2.5 $\times$ 10$^{-4}$ \\
$\bar{C}_{2,1}$              && -9.11665  $\times$ 10$^{-9}$   &  1.1 $\times$ 10$^{-9}$ &&   -1.61526  $\times$ 10$^{-8}$   &  0.4 $\times$ 10$^{-9}$ &  4.4 $\times$ 10$^{-1}$ \\
$\bar{S}_{2,1}$              &&  5.63022  $\times$ 10$^{-9}$   &  1.1 $\times$ 10$^{-9}$ &&   -1.36488  $\times$ 10$^{-8}$   &  0.4 $\times$ 10$^{-9}$ &  1.4 $\times$ 10$^{+0}$ \\
$\bar{C}_{2,2}$              &&  1.24973  $\times$ 10$^{-5}$   &  1.2 $\times$ 10$^{-9}$ &&    1.24538  $\times$ 10$^{-5}$   &  0.4 $\times$ 10$^{-9}$ & -3.5 $\times$ 10$^{-3}$ \\
$\bar{S}_{2,2}$              &&  8.52067  $\times$ 10$^{-9}$   &  1.3 $\times$ 10$^{-9}$ &&   -2.09078  $\times$ 10$^{-8}$   &  2.2 $\times$ 10$^{-9}$ &  1.4 $\times$ 10$^{+0}$ \\
$\bar{C}_{3,0}$              && -4.71444  $\times$ 10$^{-6}$   &  1.6 $\times$ 10$^{-9}$ &&   -4.76589  $\times$ 10$^{-6}$   &  1.6 $\times$ 10$^{-9}$ &  1.1 $\times$ 10$^{-2}$ \\
$\bar{C}_{4,0}$              && -5.89291  $\times$ 10$^{-6}$   &  2.4 $\times$ 10$^{-9}$ &&   -5.84911  $\times$ 10$^{-6}$   &  4.7 $\times$ 10$^{-9}$ & -7.5 $\times$ 10$^{-3}$ \\
$\bar{C}_{5,0}$              &&  2.98686  $\times$ 10$^{-7}$   &  3.5 $\times$ 10$^{-9}$ &&    2.79497  $\times$ 10$^{-7}$   &  1.2 $\times$ 10$^{-8}$ & -6.9 $\times$ 10$^{-2}$ \\
$\bar{C}_{6,0}$              &&  1.90218  $\times$ 10$^{-6}$   &  5.1 $\times$ 10$^{-9}$ &&    1.45853  $\times$ 10$^{-6}$   &  2.6 $\times$ 10$^{-8}$ &  1.5 $\times$ 10$^{-2}$ \\
\hline
\label{tab-gravSol40}
\end{tabular}
\end{table}

Our estimates of other degree-2 coefficients, ${C}_{2,1}$,
${S}_{2,1}$, and ${S}_{2,2}$, are not consistent with those of
\citet{maza14} if we use error bars corresponding to the formal
uncertainties of the fit.  However, the values are consistent if these
formal uncertainties are multiplied by a factor of 10--15 to arrive at
more realistic error bars, as suggested by \citet{maza14} and other
geodetic studies.  The difficulty in reliably estimating these
coefficients can be explained in part by the fact that they are 3--4
orders of magnitude smaller than ${C}_{2,2}$.
If Mercury were a principal axis rotator and if we
had perfect knowledge of its orientation, we would expect $C_{2,1} =
S_{2,1} = {S}_{2,2} = 0$.  Our values are close to zero (Table
\ref{tab-gravSol40}).
The misalignment between Mercury's long axis and that prescribed by
the orientation model of \citet{marg09cmda} is only $\delta \phi = 1/2
\ {\rm atan}(S_{2,2}/C_{2,2}) = 0.019^\circ$, suggesting that both the
orientation model and the recovery of ${S}_{2,2}$ are satisfactory.
Our estimate of $\delta \phi$ has a magnitude that is about 40\% of the value of \citet{maza14} and the opposite sign.  In contrast, the
IAU-defined prime meridian differs by about 0.2$^\circ$ ($\sim$8 km)
from the geophysically relevant origin of longitude, i.e., the
longitude that is defined by the principal axes of inertia and that
faces the sun at every other perihelion passage.  A spherical harmonic
expansion of the shape of Mercury reveals a long axis that is offset
by 15$^\circ$~\citep{perr15} compared to the principal axis of inertia
defined here.

Other zonal coefficients have a special significance in the context of
mission planning.  Knowledge of ${C}_{3,0}$ and ${C}_{4,0}$ can help
predict the evolution of the orbit of BepiColombo, a planned
spacecraft mission to Mercury~\citep{Genova13}.  Our estimates of
${C}_{3,0}$ and ${C}_{4,0}$ agree within 1.1\% and 0.7\% of the HgM005
values, respectively.

Table \ref{tab-gravSol40} shows that our estimate of $GM$ is also 
fully consistent with the HgM005 value.  The agreement between two
independent solutions gives confidence in the robustness of the force
and data models as well as in the overall solution strategies.
HgMUCLA40x40 and HgM005 were obtained with different software and
different approaches.  Neither solution relied on a priori information
other than the Mariner 10 results~\citep{Anderson87}.

\begin{figure}
  \noindent
\includegraphics[width=\textwidth]{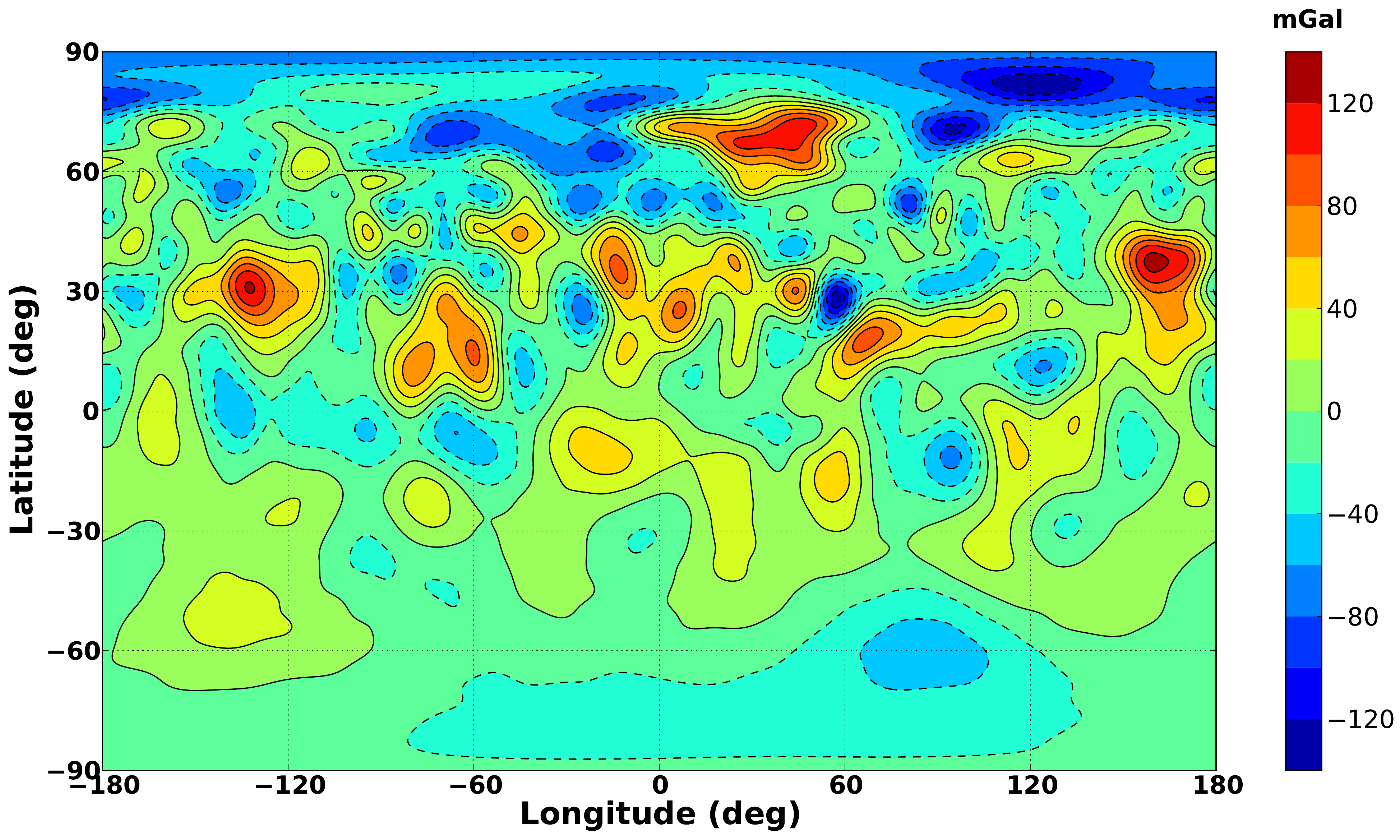}
\caption{Free-air gravity anomalies (mGal) shown on a cylindrical
  projection.  A mGal corresponds to $10^{-5} {\rm ms}^{-2}$, which is
  $\sim$1 ppm and $\sim$3 ppm of the acceleration at the surface of
  Earth and Mercury, respectively.  A large positive gravity anomaly
  is observed within the Caloris Basin plains near 160$^\circ$
  E. longitude, 37$^\circ$ latitude.}
\label{fig-anomalies}
\end{figure}

Our gravity field solution enables the calculation of free-air gravity
anomalies, which are useful in detecting internal density
inhomogeneities and in evaluating the degree of isostatic adjustment
of geological features.  Our map of free-air gravity anomalies
(Fig.~\ref{fig-anomalies}) is similar to the map derived from HgM005.
{Expansions of the geoid or free-air gravity anomalies can be used to
compute gravity-to-topography admittance ratios and evaluate the
thickness of Mercury's crust~\citep[e.g.,][]{pado15,jame15}.}

\subsection{Tidal Love Number $k_2$}
\label{loveK2}

A measurement of the tidal Love number $k_2$ is useful because it
enables us to rule out a range of interior models that are otherwise
compatible with observations of the spin and gravity of
Mercury.
The Love number can also inform us about the mechanical properties of
the mantle and the possibility of a solid FeS layer at the top of the
core~\citep{pado14jgr}.

We obtained an independent estimate of the tidal Love number $k_2$ as
part of our gravity solution HgMUCLA40x40.
We used the 0.485 $\pm$ 0.035 mean value of the theoretical estimates
\citep{pado14jgr} as an a priori value \added{and uncertainty}.
Our solution is
\begin{equation}
  k_2 = 0.464 \pm 0.023, 
  \label{eq-k2}
\end{equation}
where the adopted error bar corresponds to 10 times the formal
uncertainty of the fit. The standard
  deviation of the $k_2$ estimates across the 10 individual batch
  solutions is 0.0041, a factor of $\sim$5 smaller than our adopted
  uncertainty.  We also solved for $k_2$ with different initial
  conditions.  In a first test, we repeated our procedure with a
  variety of initial conditions ($k_2$ = 0.42, 0.45, 0.51) and found
  results consistent with our adopted solution.  In a second test, we
  started with extreme values ($k_2$ = 0 and $k_2$ = 1) and a large a
  priori uncertainty ($\sigma_{k_2} = 0.5$) in the simpler case of a
  gravity field with degree and order 20.  We found $k_2 = 0.42 \pm
  0.04$ in both instances, suggesting that we arrive at roughly the
  same $k_2$ value regardless of starting condition.  Our preferred
  value is the HgMUCLA40x40 solution (Equation (\ref{eq-k2})), which
  is compatible with the value of \citet{maza14} and the computed
  value of \citet{pado14jgr}.  We emphasize that a reliable
  measurement of $k_2$ from MESSENGER tracking data is challenging
  because the spacecraft is in a highly eccentric orbit and is also
  subject to substantial non-gravitational perturbations.  In the
  course of our gravity solutions with different strategies and
  parameters, we encountered best-fit values for $k_2$ in the range
  0.420--0.465.

\subsection{Spin Axis Orientation}
\label{spinAxis}
A priori values for the spin state of Mercury were
based on the libration model of \citet{marg09cmda} with small
adjustments to the libration amplitude and spin orientation values as
recommended by \citet{marg12jgr} (Section~\ref{gfm}).  Although we did
not attempt to fit for the {spin rate or} libration amplitude at this
time, our model can be expanded to perform such a fit in the
future. {To do so, we would express Mercury's rotational phase as a
  trigonometric series and solve for the series coefficients.}

\begin{figure}[h]
  \centering
\includegraphics[width=0.7\textwidth]{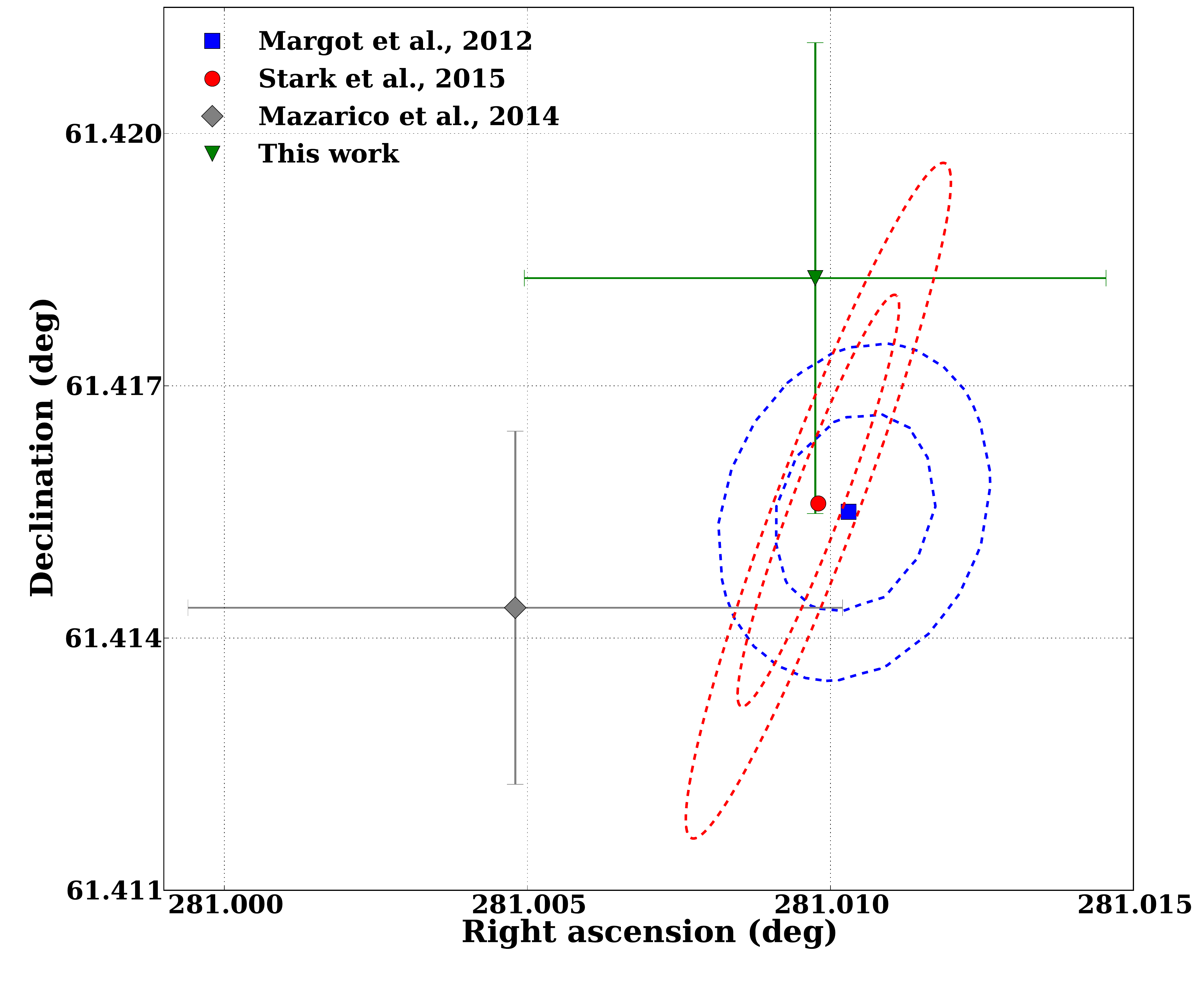}
\caption{Spin axis orientation solutions: Earth-based radar (blue),
  altimeter + DEMs (red), HgM005 gravity (gray), and HgMUCLA40x40 gravity (green, this work).}
\label{fig-SpinAxis40}
\end{figure}

{The recovery of the spin axis orientation exhibited good
  consistency across the 10 individual batch solutions.  The standard
  deviations of the right ascension and declination estimates across
  the 10 batches are 0.00089$^\circ$ and 0.00063$^\circ$, respectively.}
 Our final, weighted mean estimate of the spin axis orientation at epoch J2000.0 is
\begin{equation}
{\rm RA} = 281.00975^\circ \pm 0.0048^\circ,
\end{equation}
\begin{equation}
{\rm DEC} = 61.41828^\circ \pm 0.0028^\circ,
\end{equation}
where we adopted error bars corresponding to 10 times the formal
uncertainties of the fit.  Our gravity-based estimate of the spin axis
orientation is within 10 arcseconds of the independent crust-based
estimates of \citet{marg12jgr} and \citet{star15grl}
(Figure~\ref{fig-SpinAxis40}).  The convergence of all 3 values is
important because of the profound impact of Mercury's obliquity on the
determination of the moment of inertia and, consequently, on the
construction of accurate interior models.  Our solution is 16
arcseconds away from and only marginally consistent with the
gravity-based estimate of \citet{maza14}.  The lack of a better
agreement between HgMUCLA40x40 and HgM005 for the spin pole parameters
despite a generally excellent agreement in the gravity coefficients is
worth noting.  We speculate that the difference might be due to our
use of an improved ephemeris, DE432, compared to \citet{maza14}'s use
of DE423, or to our inclusion of range data, which they may not have included.

\subsection{Earth-Mercury Distance Error}
\label{merEph}
\begin{figure}[h]
\centering
\includegraphics[width=\textwidth]{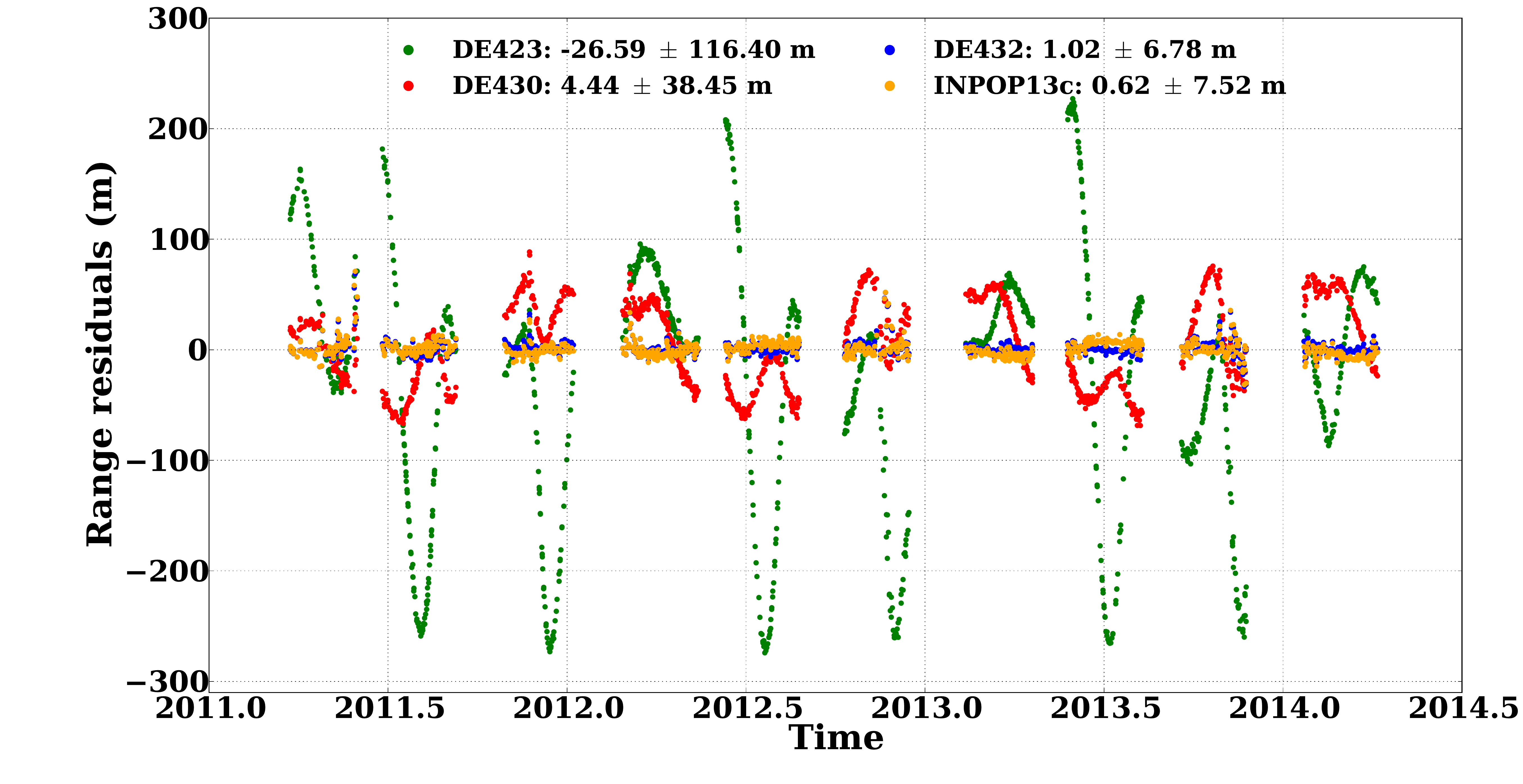}
\caption{One-way range residuals obtained with the DE423, DE430,
  DE432, and INPOP13c planetary ephemerides.  Each point represents
  the mean residual corresponding to an individual arc.  The legend
  shows the mean and standard deviation of the residuals for each data
  set.  Gaps correspond to solar conjunction periods (SPE $< 40^\circ$). }
\label{fig-ephBias}
\end{figure}

{
Range measurements to the MESSENGER spacecraft currently provide the
best way of improving the knowledge of Mercury's position~\citep{verm14,maza14,fienga15}.
We used DE432~\citep{FolknerDE432} as the nominal planetary ephemeris
when deriving the HgMUCLA40X40 gravity solution.  We computed
MESSENGER range residuals using this ephemeris and compared them to
the residuals obtained with DE423~\citep{FolknerDE423},
DE430~\citep{FolknerDE430-31}, and INPOP13c~\citep{fienga15}.  Figure
\ref{fig-ephBias} illustrates the remaining Earth-Mercury distance
error in the various ephemeris versions.  DE423 includes range data
obtained during MESSENGER flybys but not during the orbital phase of
the mission and retains errors in Earth-Mercury distance at the
$\sim$100 m level.  DE430 includes range data from the first 6 months
of the orbital phase.  Both DE432 and INPOP13c include several years
of range data from the orbital phase of the mission.  Although the
mean of INPOP13c residuals is lower than that of DE432 due to
fortuitous cancellation of positive and negative values, close
inspection reveals that systematic patterns in the residuals are more
pronounced in the INPOP13c solution.  Thanks to MESSENGER data, the
knowledge of the Earth-Mercury distance in the 2011-2014 interval is
now at the $\sim$7~m level.  MESSENGER range and Doppler data can be
combined with similar data from other missions as well as radar and
optical astrometry of all planets and asteroids to further improve
solar system ephemerides.}

\section{Discussion}

{Our recovery of low-degree coefficients in the spherical harmonic
expansion of the gravity field is generally in good agreement with
previous results (Table~\ref{tab-gravSol40}).  }

The gravity-based solution of \citet{maza14} for Mercury's spin axis
orientation was found to be marginally consistent with the Earth-based
radar and topography-based solutions (Figure~\ref{fig-SpinAxis40}),
but also sufficiently different to raise questions about the
possibility of a discrepancy.  A discrepancy would indicate either an
error in one of the determinations, or a real difference between the
orientations about which the gravity field and the crust rotate.
Although the latter prospect is intriguing, we find no convincing
evidence for a discrepancy.  Our spin pole estimate is close ($10$
arcseconds) to the crust-based spin pole measurements but in a
different direction than that identified by \citet{maza14}.  The most
plausible conclusion is that both gravity-based estimates are
marginally consistent with the crust-based estimates and that the
gravity field and crust rotate about nearly the same axis.

Our recovery of Mercury's spin axis orientation from gravity data
alone suggests that gravity-based methods can be applied successfully
to reach a similar goal at other bodies. 
Other examples include the gravity-based spin axis orientation of
  Venus~\citep{kono99} which is on the edge of the uncertainty region
  of the crust-based estimate~\citep{davi92}.  The situation at Mars
  is not directly comparable due to the availability of data from
  multiple long-lived landers.  The gravity-based spin axis
  orientation of Vesta~\citep{kono14} is in good agreement with the
  crust-based estimate~\citep{russ12}.

\begin{figure}[h]
  	\centering
\noindent
\includegraphics[width=30pc]{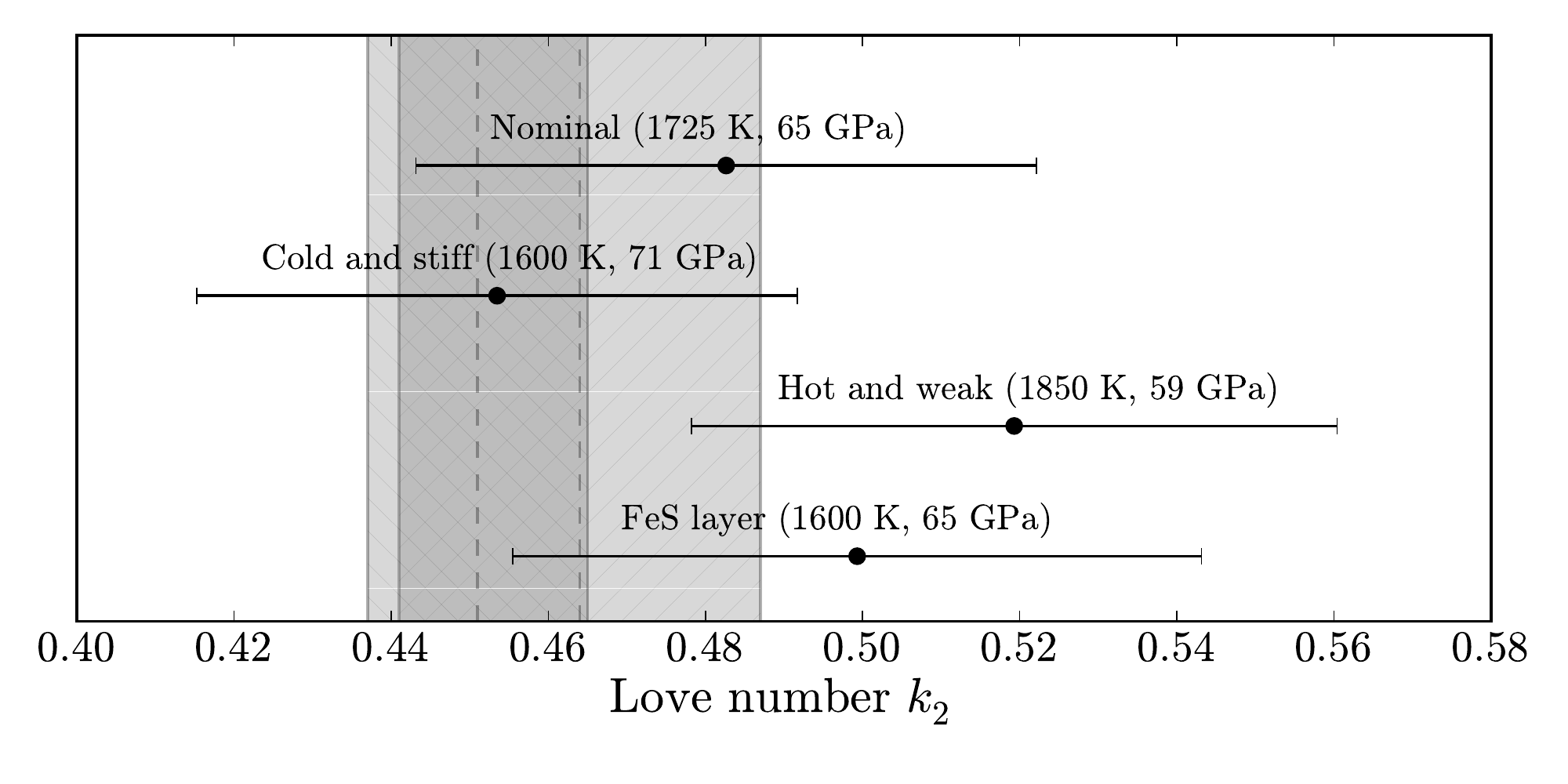}
           \caption{Comparison of measured and calculated values of
             the tidal Love number $k_2$.  The black solid dots with
             error bars represent 4 classes of models investigated by 
             \citet{pado14jgr} with different assumptions about the
             basal temperature and rigidity of the outer shell.  The
             vertical dashed lines correspond to our measured $k_2$
             value and that of \citet{maza14}.  The descending hatch
             pattern represents the one-standard-deviation
             uncertainties of \citet{maza14}. Our adopted
             uncertainties are shown with the ascending hatch
             pattern.}
\label{fig-k2}  
\end{figure}

Our estimate of Mercury's tidal Love number ($k_2 = 0.464 \pm 0.023$)
is larger than an existing determination ($k_2 = 0.451 \pm 0.014$,
\citep{maza14}) and admits a wider range of interior models.
\citet{pado14jgr} simulated the tidal response of Mercury on the basis
of tens of thousands of interior models with different physical
properties.  Two parameters that strongly influence the tidal response
are the temperature at the base of the mantle and the rigidity of the
mantle (Fig.~\ref{fig-k2}).  \citet{pado14jgr} showed that the $k_2$ estimate of
\citet{maza14} favors a cold and stiff mantle (basal temperature of
1600 K, rigidity of 71 GPa) with no FeS layer.  Our Love number
determination is compatible with a substantially larger set of models,
including \citet{pado14jgr}'s nominal class of models with a hotter
and weaker mantle (basal temperature of 1725 K, rigidity of 65 GPa).
It is also compatible with models that include an FeS layer at the
bottom of the mantle, and even with a small fraction of hot and weak
models (basal temperature of 1850 K, rigidity of 59 GPa).

\section{Conclusions}
{We analyzed over three years of MESSENGER radio tracking data.
  We estimated parameters that describe a spherical harmonic expansion
  to degree and order 40 of Mercury's gravity field as well as the
  tidal Love number and spin axis orientation.  Our solution for
  Mercury's mass and gravity field is in excellent agreement with
  previous estimates.  In particular, recovery of the $C_{2,0}$ and
  $C_{2,2}$ coefficients gives additional confidence in prior
  inferences about Mercury's moment of inertia and interior structure.
  Our estimate of the tidal Love number is larger than
  \citet{maza14}'s estimate, which favored interior models with a cold
  and stiff mantle.  Our estimate is compatible with a wider range of
  interior models, including models with a hotter and weaker mantle.
  Our spin state solution is consistent with previous measurements of
  the orientation of the crust but only marginally consistent with
  \citet{maza14}'s estimate.  We hypothesize that the difference is
  related to our use of an improved ephemeris with range residuals at
  the $\sim$7 m level (vs. $\sim$116 m) or to our inclusion of range
  data in the solution.  Finally, we confirmed that the Earth-Mercury
  distance in the 2011-2014 interval is now known to better than 10
  m.}

\acknowledgments
We thank anonymous reviewers for suggestions that improved the manuscript.
This work was enabled in part by the Mission Operations and Navigation
Toolkit Environment (MONTE).  MONTE is developed at the Jet Propulsion
Laboratory, which is operated by Caltech under contract with NASA.
The archive locations for the data used in this work are listed in the
text and references.  The HgMUCLA40x40 coefficient values and
uncertainties are available from the University of California archive
escholarship at http://escholarship.org/uc/item/74c8x8mj.
AKV and JLM were supported in part by the NASA Planetary Astronomy
program under grant NNX12AG34G and by the MESSENGER Participating
Scientist program under grant NNX09AR45G.


\listofchanges

\end{document}